\documentclass[12pt]{article}
\usepackage{graphicx}
\usepackage{graphicx}

\begin{document}

\title{ Stokes' Second Flow Problem   in a High Frequency  Limit: Application to Nanomechanical -Resonators.
\\}

\author {Victor Yakhot and Carlos Colosqui, 
\\
Department of Aerospace and Mechanical Engineering, \\
Boston University, Boston, MA 02215 \\}

\maketitle

\begin{abstract}
\noindent    Solving the Boltzmann - BGK equation, we investigate
a flow generated by an infinite plate oscillating with frequency $\omega$.  Geometrical
simplicity of the  problem allows a solution in the entire range of  dimensionless 
frequency variation $0\leq \omega \tau\leq \infty$, where $\tau$ is a  properly defined  relaxation time.  
A  transition from viscoelastic behavior of Newtonian fluid ($\omega\tau\rightarrow 0$)  to purely elastic dynamics in the limit $\omega\tau\rightarrow \infty$ is discovered.   
 The relation of the derived solutions to microfluidics  (high-frequency micro-resonators) is demonstrated on an example of a  "plane oscillator".  
\end{abstract}

\vspace{0.3in}


\noindent {\it Introduction.}  During last two centuries,   Newtonian fluid approximation  was remarkably
successful 
in explaining a wide  variety of natural phenomena ranging from flows in pipes, 
channels and boundary layers  to the  recently discovered processes   in meteorology, aerodynamics
, MHD and cosmology.  With advent of powerful computers and development of effective
numerical methods,  Newtonian hydrodynamics   remains  at the foundation  
of   various design tools widely used in mechanical and civil engineering.  Since
technology of the past mainly dealt with large ( macroscopic)  systems varying on
the length/ time-scales $L$ and $T$,  the Newtonian fluid approximation, typically 
defined by  the smallness of  Knudsen  and Weisenberg numbers $Kn=\lambda/L\ll 1$
and $Wi=\tau/T\approx \frac{\lambda}{L}\frac{u}{c}=KnMa\ll
1$,  was accurate enough.  The length and time-scales $\lambda$ and $\tau\approx\lambda/c$,  
are the  mean -free path and relaxation time,    respectively,
.

\noindent With recent  rapid developments in nanotechnology and bioengineering, 
quantitative description of high-frequency oscillating microflows,  i. e. flows  where    Newtonian approximation
breaks down,  became an important and urgent task from both basic  and applied science
viewpoints. 
Modern micro (nano)- electromechanical devices (MEMS and NEMS),   operating  in 
the high- frequency range up to  $\omega=O(10^{8}-10^{9}Hz)$ ($Wi=\omega
\tau \geq 1$),  can lead to  small  biological mass detection,  subatomic microscopy,
viscometry and other applications [1]-[3].   The manufactured devices are so small  and sensitive that  adsorption   of even tiny particles on their surfaces leads to   detectable   response in  the resonator frequency peak   (shift) and  quality  ($Q$) factor,  thus  enabling these revolutionary  applications.  Since  both frequency  shift and  width of the  resonance peak,  depend upon properties
of the resonator -generated flows,  the microresonators  may  serve as sensors  
enabling  accurate investigation  of  fundamental processes  in  microflows.
 
\noindent  Often, oscillating  flows are  a  source of   new and unexpected phenomena.
For example,  
in a recent study of  the high-frequency elctro-magnetic -  field - driven nano 
-resonators  ($h\times w\times L\approx 0.2\times\times 0.7\times 10\mu m$)  Ekinci
and Karabacak ,  
 observed  a  transition  in the frequency dependence   of inverse quality factor  from  $1/Q\approx \gamma/\omega\propto
1/\sqrt{\omega}$,  expected in the hydrodynamic limit,   to $1/Q\propto 1/\omega$ in the high-frequency (kinetic)  limit [3].   Since  parameter $\gamma\propto W$ is proportional  to the energy  
dissipation rate ($W$)  into a surrounding gas,  the discovered effect  points to
the frequency- independent dissipation rate $W$.  No quantitative theory  describing this transition  has been developed.

\noindent Recently,  Shan et.  al.  performed a  detailed numerical  investigation of evolution  of  the initially prepared viscous shear layer defined by  a unidimensional velocity field ${\bf u}(y,0)=u_{0}\cos( k_{0}y){\bf i}$, where ${\bf i}$ is the unit vector in the $x$-direction. A  novel transition from purely viscous decay to the viscoelastic dynamics  was discovered for $\tau\nu k_{0}^{2}\geq 1$ [4].

\noindent In this paper we investigate  deviations from  Newtonian hydrodynamics,
on  an  example of   classic " Stokes' Second "  problem of a flow generated
by an infinite  solid plate oscillating  along  $x$-axis with velocity $u(y=0,t)=U\cos
\omega t$ [5].    First we use the
hydrodynamic approximation,  derived  from  kinetic equation by Chen et. al..[6] 
in the limit $\omega\tau  \ll 1$.  Then,   derivation of  equations valid in an infinite 
 range $0<\omega\tau<\infty $   is presented  and it is shown  that 
in the  high-frequency limit $\omega\tau \rightarrow \infty$,  the dissipation rate
$W\rightarrow const$.  To apply  the derived solution to experimental data on nano-resonators, we introduce a model system of plane oscillator  and  demonstrate the transition in the frequency dependence of the quality factor,  in a quantitative agreement with experimental data of  in Refs. [4], [5].




\noindent {\it  The Boltzmann-BGK equation}.   Interested in the rapidly oscillating flows , where  the Navier-Stokes equations break down,  we consider the kinetic equation in the relaxation time approximation (RTA):

\begin{equation}
\frac{\partial f}{\partial t}+{\bf v}\cdot \nabla 
f={\cal C}
\end{equation}

\noindent  In accord with Boltzmann's  H-theorem , the initially non-equilibrium gas  must monotonically relax to thermodynamic equilibrium.  This leads  to the relaxation time  anzatz,  qualitatively satisfying this requirement:
\begin{equation} 
{\cal C} 
\approx - \frac {f - f^{eq}} {\tau (v)}
\label{bgk}
\end{equation}        

\noindent  In thermodynamic  equilibrium  the left  side of kinetic  equation is
equal to zero and the remaining equation ${\cal C}=0$ has a solution:
$
f^{eq}=constant\times e^{-\frac{E({\bf p,r})}{\theta}}
$  with the temperature $\theta=const$.   
In the mean -field approximation, valid close to equilibrium where all gradients are small,  one replaces   the variable  $v$ by  $\overline
{v}\propto \sqrt{T/m}$.  Given  the  mean-free path $\lambda$,   we have an estimate 
 $\tau=\lambda/\overline{v}\approx \lambda/c=const$  where $c$ is the speed of sound.
 In general,  the relaxation time can be a non-trivial function of velocity gradients,  position in the flow, external fields etc  [7]. The equation (1)-(2) is the celebrated Boltzmann BGK equation widely used for both theoretical and numerical (Lattice Boltzmann Method) studies of non-equilibrium fluids [7]. This equation,  with 
 
\begin{equation}
f^{eq}=\frac{\rho}{(2\pi\theta)^{\frac{3}{2}}}\exp(-\frac{({\bf v-u})^{2}}{2\theta})
\end{equation} 
 
 \noindent  often considered as a generic  equation of fluid dynamics, is valid  when   velocity gradients are not large.  It will be shown below that, due to a particular geometry  studied in this paper,   the largest magnitude of velocity gradient $|\partial_{y} u(0,t)| < U / \delta$ with $\delta\rightarrow \infty$  when $\omega\tau\rightarrow \infty$. Therefore, when the amplitude $U$ is small enough, the typical gradient is small. Moreover,  this feature is responsible for the purely elastic response and total disappearence of viscosity from the problem.  In turbulence theory this effect is called rapid distortioin (RD) limit.   
  
\noindent  {\it Chapman-Enskog expansion of BGK. Chen et.al (2003).} Multiplying (1),(2) by $v$ and integrating  over ${\bf v}$ gives:

\begin{equation}
\frac{\partial u_{i}}{\partial t}+u_{j}\frac{\partial u_{i}}{\partial x_{j}}+\frac{1}{\rho}\frac{\partial}{\partial x_{i}} \rho \sigma_{ij}=0
\end{equation}
\noindent where the stress tensor, written for $i\neq j$ is:

\begin{equation}
 \sigma_{ij}=\overline{(v_{i}-u_{i})(v_{j}-u_{j})}\equiv \int d{\bf v}(v_{i}-u_{i})(v_{j}-u_{j})f({\bf v, x},t)
\end{equation}

\noindent Usually, evaluation of the stress tensor is  a difficult task. The simplified  equation  (1)-(2)  for a  single-particle distribution
function 
allows   calculation of nonlinear contributions to the  momentum stress tensor $\sigma_{ij}=\overline{(v_{i}-u_{i})(v_{j}-u_{j})}-\frac{1}{d}\overline{v^{2}}\delta_{ij}$.  
In a remarkable paper,  Chen et. al. [6],  based on thequation (1)-(2),    formulated the Chapman-Enskog expansion in powers on dimensionless relaxation time  $\epsilon=\tau\partial _{x_{i}}u_{j}$.  In this formulation:

\begin{equation}
\nabla=\epsilon \nabla_{1}
\end{equation}

\begin{equation}
\frac{\partial}{\partial t}= \epsilon \frac{\partial}{\partial t_{0}}+\epsilon^{2}  \frac{\partial}{\partial t_{1}}+\epsilon^{3}\frac{\partial}{\partial t_{2}}+\cdot\cdot\cdot
\end{equation}

\noindent and the pdf is expanded in powers of $\epsilon$ as:
\begin{equation}
f=f^{(0)}+\epsilon f^{(1)}+\epsilon f^{(2)}+\cdot\cdot\cdot
\end{equation}

\noindent Substituting these expressions into (1),(2) and equating the terms of the same order in $\epsilon$  results in  $f^{(0)}=f^{eq}$ given by (3),  and 

\begin{equation}
(\frac{\partial}{\partial t_{0}}+{\bf v}\cdot \nabla)f^{(0)}=-\frac{f^{(1)}}{\tau}
\end{equation}

\begin{equation}
(\frac{\partial}{\partial t_{0}}+{\bf v}\cdot \nabla)f^{(1)}+\frac{\partial}{\partial t_{1}}f^{(0)}=-\frac{f^{(2)}}{\tau}
\end{equation}

\noindent etc. The mean of any fluctuating variable is then calculated as :

\begin{equation}
\overline{A({\bf v})}=A^{(0)}+A^{(1)}+A^{(2)}+\cdot\cdot\cdot
\end{equation}

\noindent where

$$A^{(n)}=\int f^{(n)} A({\bf v})  {\bf v}$$

\noindent  To illustrate the main results and simplify notation in what follows we set the temperature $\theta=const$. The calculation of Chen (2003) gives:

\begin{equation}
f^{(1)}\approx -\frac{\tau}{\theta}f^{(0}S_{ij}[(v_{i}-u_{i})(v_{j}-u_{j})-\frac{({\bf v-u})^{2}}{d}\delta_{ij}]
\end{equation}

\noindent and 

\begin{equation}
f^{(2)}=-2\tau^{2}f^{(0)}[(v_{i}-u_{i})\frac{\partial}{\partial x_{j}}(S_{ij}-\frac{1}{d}\nabla\cdot{\bf u}\delta_{ij})+O((v_{p}-u_{p})(v_{q}-u_{q})S_{pi}S_{qi})]
\end{equation}

\noindent It is important to stress that the last contribution to the right side of (13) involves even
 powers of $v-u$ and various products of $S_{ij}$.  The result is (Chen (2003)):
 
\begin{eqnarray}
\sigma_{ij} &=& 
2\tau\theta S_{ij} - 2\tau\theta (\partial_t + {\bf u} \cdot
\nabla )(\tau S_{ij})
\nonumber \\
& & - 4 \tau^2 \theta 
[S_{ik} S_{kj} - \frac {1} {d} \delta_{ij}
S_{kl} S_{kl} ] 
+ 2 \tau^2 \theta [S_{ik} \Omega_{kj} + S_{jk} \Omega_{ki}]
\label{sed}
\end{eqnarray}
where  the rate of strain $S_{ij}=\frac{1}{2}[\frac{\partial u_{i}}{\partial x_{j}}+\frac{\partial u_{j}}{\partial x_{i}}]$ and the vorticity tensor is defined as,
$ \Omega_{ij} = \frac {1} {2} [\frac {\partial
u_i} {\partial x_j} 
- \frac {\partial u_j} {\partial x_i}] $. 
 
\noindent The first  term in  the right side of (14),   resulting from the first order Chapman-Enskog
(CE) expansion,
corresponds to the familiar Navier-Stokes equations for {\it Newtonian fluid } and 
the non-linear ({\it non - Newtonian})   corrections,  given by the remaining terms,  are 
generated in the next ,  second,  order. The constitutive relation (14) is quite  complex and, in general,  can be
attacked by numerical methods only.  However, it is  greatly simplified
in  an important  class of simple unidirectional flows.

  \noindent  {\it  Stokes'  Second  Problem}  is  formulated as follows:
 The flow of a fluid filling  half -space $0\leq y$ is generated by a   solid plate
at $y=0$ moving along  the $x$-axis with 
 velocity $U_{p}(0,t)=U\cos \omega t$. Since velocity components in
$y$ and $z$-directions are equal to zero,
we have to solve the  equations  for the $x$-component of  velocity  field $u(y,t)$ only. Due to geometry of the problem  

\begin{equation}
u_{x}\equiv u=u(y,t);  ~~ u_{y}\equiv v =0; ~~{\bf u}\cdot \nabla =0;~~\partial_{x}=0 
\end{equation}
Thus, since $i\neq j$, we are interested in $\sigma_{x,y}$.  In this case,  the equation  of motion corresponding to  the stress tensor  (14) is very simple:
 
 \begin{equation}
 \frac{\partial u}{\partial t}=\nu(1-\tau\frac{\partial
}{\partial t})\frac{\partial^{2} u}{\partial
y^{2}}
 \end{equation}
\noindent  In the limit $\omega\tau\rightarrow 0$, this equation is to be solved subject to the  no-slip boundary condition,
$
u(0,t)=U\cos\omega t;~~\lim_{y\rightarrow \infty} u(y,t)=0. 
$
\noindent 
Seeking a  solution, satisfying the  boundary condition  at
$y\rightarrow\infty$,   as:
$
u={\cal R}( \phi(y)e^{-i\omega t})
$
\noindent  gives: $
\phi =Be^{-\frac{y}{\delta}}
$
and in the low frequency limit  $Wi=\tau\omega\ll1$:



\begin{equation}
u(y,t)=Uexp(-\sqrt{\frac{\omega}{2\nu}}(1-\frac{\tau\omega}{2})y)\cos
(\omega t-\sqrt{\frac{\omega}{2\nu}}(1+\frac{\tau\omega}{2})y)
\end{equation} 
 
\noindent  In a classic case $\omega\tau=0$,  the flow is characterized by a
single scale 
$\sqrt{2\nu/\omega}$ describing both  "penetration depth"
and  wave-length of transverse waves radiated  by the oscillating  plate. We 
can see that the non-newtonian  $O(\tau \omega)$ contribution leads  to formation
of two different length-scales:   the $O(\tau\omega)$ increasing   penetration depth
and by the wave-length decreasing by the same magnitude.  This is a qualitatively
new feature of a non-newtonian flow.  Now we calculate the dissipation rate. The
force acting on the unit area of the wall at $y=0$ :

\begin{equation}
\mu(1-\tau\frac{\partial}{\partial t})\frac{\partial u(0,t)}{\partial
y}=
-U\sqrt{\omega\mu\rho}[\cos(\omega t+\frac{\pi}{4})+\frac{\tau\omega}{2}\cos(\omega
t-\frac{\pi}{4})]
\end{equation}  

\noindent  is phase-shifted  relative to velocity field $u(0,t)=U\cos \omega
t$. This  result differs from 
its  classic Newtonian counterpart by an $O(\tau\omega/2)$ shift. The mean energy
dissipated per unit time per unit are of the plate  is calculated if  we multiply
(18) by $-U\cos \omega t$ and integrate over one cycle with the result:

\begin{equation}
W(\tau,\omega)=\frac{U^{2}}{2}\sqrt{\frac{\omega\mu\rho}{2}}(1+\frac{\tau
\omega}{2})>W(0,\omega)
\end{equation}

\noindent In a simple gas close to thermodynamic equilibrium, the mean-free path $\lambda\approx \frac{1}{\sqrt{2}\Sigma
n}\propto \frac{k_{B}T}{d^{2}p}$
where $\Sigma\propto d^{2}$ is a scattering cross-section of the molecules
and $d$ is a typical scale of intermolecular interaction.      \\

\noindent {\it Large deviations from Newtonian  fluid mechanics.} Here we somewhat modify the procedure developed by Chen (2003).    It is clear that as $\tau\rightarrow 0$, the expansion gives the classic Stokes results for the flow of oscillating plate. Moreover, in the limit $\tau\omega\ll 1$, the second -order    in ${\bf S}$ correction to (14) disappears  due to the symmetries of the problem defined by (7).
The expression (14) includes a time derivative, which for the oscillating flow we are interested in this paper,  introduces   an additional dimensionless parameter $Wi-\tau\omega$ into expansion. Therefore, the CE expansion is in fact expansion in two dimensionless parameters $Kn=\tau \partial_{y}u$ and $Wi$. As  will be shown below, in the simple-geometry oscillating flows,  these parameters are quite different : as  $\tau\omega\rightarrow \infty$, the second parameter $Kn\rightarrow 0$. Thus,  while  neglecting the small,   $O(\nabla^{2n}u)$  with $n>1$, 
 "Burnett contributions",    we will attempt to sum up the entire series in $Wi=\tau\omega$ .  The Boltzmann equation is:

\begin{equation}
\frac{\partial f}{\partial t}+v_{\alpha}\frac{\partial
f}{\partial x_{\alpha}}+\frac{f}{\tau}=\frac{f^{e}}{\tau}
\end{equation}

\noindent Multiplying (20) by $(v_{i}-u_{i})(v_{j}-u_{j})$ and integrating over $\bf v$, we derive the equation for the stress tensor  ( $i\neq j$):  :

\begin{equation}
\frac{\partial \sigma_{ij}}{\partial t}+\frac{\sigma_{ij}}{\tau}=-\frac{1}{\rho}\int
d{\bf v} (v_{i}-u_{i})(v_{j}-u_{j})v_{\alpha}\frac{\partial
f}{\partial x_{\alpha}}
\end{equation}
\noindent or

\begin{equation}
\frac{\partial \sigma_{ij}}{\partial t}+{\bf u}\cdot  \nabla \sigma_{ij}+\frac{\sigma_{ij}}{\tau}=-2\sigma_{i,\alpha}\frac{\partial u_{j}}{\partial x_{\alpha}}-\frac{\partial }{\partial x_{\alpha}}\overline{(v_{\alpha}-u_{\alpha})(v_{i}-u_{i})(v_{j}-u_{j})}-\sigma_{ij}\nabla\cdot{\bf u}
\end{equation}

\noindent Due to the basic symmetries of the problem (7), this equation is simplified:

\begin{equation}
\frac{\partial \sigma_{ij}}{\partial t}+\frac{\sigma_{ij}}{\tau}=-2\sigma_{i,\alpha}\frac{\partial u_{j}}{\partial x_{\alpha}}
-\frac{\partial }{\partial x_{\alpha}}\overline{(v_{\alpha}-u_{\alpha})(v_{i}-u_{i})(v_{j}-u_{j})}
\end{equation}

\noindent In the unidirectional flow, we are interested in only $\partial_{y}u\neq 0$ and therefore,

\begin{equation}
\sigma_{i,\alpha}\partial_{x_{\alpha}}u=\sigma_{yy}\partial_{y}u
\end{equation}

\noindent The remaining term in the right side of equation (15) can also be simplified leading to:

\begin{equation}
\frac{\partial \sigma_{y,x}}{\partial t}+\frac{\sigma_{y,x}}{\tau}=-2\theta\frac{\partial u}{\partial y}
-\frac{\partial }{\partial y}\overline{v_{y}^{2}(v_{x}-u)}
\end{equation}

\noindent where $\theta=\sigma_{y,y}=\overline{v_{y}^{2}}=\frac{({\bf v-u})^{2}}{3}$. 
This equation   is formally exact. 

\noindent Our goal now is to evaluate 

\begin{equation}
\Sigma_{ij}=\frac{\partial }{\partial x_{\alpha}}\overline{(v_{\alpha}-u_{\alpha})(v_{i}-u_{i})(v_{j}-u_{j})}=\Sigma_{ij}^{(0)}+\Sigma_{ij}^{(1)}+\Sigma_{ij}^{(2)}+\cdot\cdot\cdot
\end{equation}
\noindent This can be done readily with relations (3),  (11)-(13) derived in Chen (2003). We see that $\Sigma_{ij}^{(0)}=0$ and $\Sigma_{ij}^{(1)}=0$.  Evaluated on $f^{(2)}$ from (13)

\begin{equation}
\Sigma_{xy}^{(2)}\approx \tau^{2}\frac{\partial }{\partial y}\overline{v_{y}^{2}(v_{x}-u_{x})^{2}}|_{0}\frac{\partial^{2} u}{\partial y^{2}}\approx \tau^{2}\theta^{2}\frac{\partial ^{3} u}{\partial y^{3}}
\end{equation}

\noindent where  $\overline{A}|_{0}=\int f^{0}({\bf v})A({\bf v})d{\bf v}$. It will be shown a'posteriory that in both limits $\tau\omega\rightarrow 0$ and $\tau\omega\rightarrow \infty$, the second order contribution 
\begin{equation}
\frac{\Sigma_{ij}^{(2)}}{\theta\partial_{x}u }\approx \frac{\tau^{2}\theta}{\delta^{2}}\rightarrow 0
\end{equation}
\noindent where $\delta$ is the width of the viscous layer. 

\noindent As we saw, the high powers of ${\bf S}$ disappear from the second-order relation (14).  
 Let us see  that this is a general phenomenon valid to all orders. 
 Let us consider a general  $n^{th}$ -order 
contribution to the stress tensor $\sigma_{x,y}$:

\begin{equation}
\frac{\partial u_{\alpha_{1}}}{\partial x_{\alpha_{3}}}\frac{\partial u_{\alpha_{2}}}{\partial x_{y}}\frac{\partial u_{\alpha_{3}}}{\partial x_{\alpha_{n-1}}}\cdot\cdot\cdot \frac{\partial u_{x}}{\partial x_{\alpha_{n}}}\cdot\cdot\cdot \frac{\partial u_{\alpha_{n-1}}}{\partial x_{\alpha_{2}}}\frac{\partial u_{\alpha_{n}}}{\partial x_{\alpha_{1}}}
\end{equation}

\noindent where the summation is carried out over the randomly distributed Greek subscripts. By the symmetry

\begin{equation}
\frac{\partial u_{x}}{\partial x_{\alpha_{n}}}\frac{\partial u_{\alpha_{n}}}{\partial x_{\alpha_{1}}}=\frac{\partial u}{\partial y}\frac{\partial v}{\partial x_{\alpha_{n}}}=0
\end{equation}

\noindent because $v=0$.  
Denoting $\sigma_{x,y}\equiv \sigma$ we, based on the above considerations obtain equation valid in both  low and large-frequency limits:

\begin{equation}
\frac{\partial u}{\partial t}=-\frac{\partial \sigma}{\partial y}; \hspace{0.5cm}
\tau\frac{\partial \sigma}{\partial t}+\sigma=-\nu\frac{\partial
u}{\partial y}
\end{equation}

 \noindent Comparing this result with the equation (16), we conclude that  constitutive equation (31) is
a resummation of the Chapman-Enskog expansion used in the low-order derivation of
(14).  Indeed, the Fourier-transform of the  second equation in (31) 
 is equal to 
 $
(1+i\omega\tau) \sigma(y,\omega)=\nu\frac{du}{dy}
$  which in the limit $\omega\tau\rightarrow 0$, coinsides with (16).  The
equations (31)   give:
 
 \begin{equation}
 \tau\frac{\partial^{2} u}{\partial t^{2}}+\frac{\partial
u}{\partial t}=\nu\frac{\partial^{2} u}{\partial
y^{2}}
 \end{equation}

\noindent  {\it Boundary conditions.}  Interested  in the limit $\omega\tau>1$,  we have to be careful with the choice of boundary conditions (BC). The BC  used in this  work  is $u(0,t)=U(\omega\tau)\cos \omega t=\alpha(\omega\tau)U\cos \omega t$. The "slip factor $0<\alpha(\omega\tau)\leq 1$ has been recently investigated in detailed numerical simulations where is was shown that  for $\omega\tau\leq 1$, $\alpha(\omega\tau)\approx 1$ and it rapidly decreases to $\alpha(\omega\tau)\approx 0.3-0.2$ for $1<\omega\tau \leq 100$ [9].  Dealing with  the linear equation,  to remove the slip factor from consideration, we can, in principle,  introduce the normalized velocity field $u/U(\omega\tau)$, solve equations  and then recover   $u(y,t)$.  Below, we use the no-slip boundary condition.

\noindent {\it solution.} The solution to equation (32) is found readily: 
$
u={\cal R}[e^{-i\omega t} e^{-\frac{y}{\delta}}]
$
 with $
\frac{1}{\delta}=-(1-i)\sqrt{\frac{\omega}{2\nu}}\sqrt{1-i\omega\tau}
$.
and:

\begin{equation}
u=U e^{-\frac{y}{\delta_{-}}}\cos(\omega
t-\frac{y}{\delta_{+}})
\end{equation}
\noindent where

\begin{equation}
1/\delta_{\pm}=(1+\omega^{2}\tau^{2})^{\frac{1}{4}}\sqrt{\frac{\omega}{2\nu}}[\cos(\frac{1}{2}\tan^{-1}(\omega\tau))\pm
\sin(\frac{1}{2}\tan^{-1}( \omega\tau))]
\end{equation}
\noindent In the limit $\omega\tau\rightarrow 0$, this solution tends to the expression
(17) derived above.
As we see,   the non-Newtonian Second Stokes problem can be characterized by  two
different length-scales,  penetration length  $\delta_{-}$ and the wave-length
$\delta_{+}$.  In the limit $\omega\tau\rightarrow \infty$, the penetration
length tends to infinity and the dominating 
dissipation  mechanism is  the wave generation. (This limits the velocity gradients in this problem). To calculate the force acting on the 
unit area of the plate,  we notice that in accord with the differential equations  
(31)

\begin{equation}
\sigma(0,t)=\rho \nu \exp(-\frac{t}{\tau})\int_{-\infty}^{t}
\frac{\partial u(0,\lambda)}{\partial y}\exp(\frac{\lambda}{\tau})d\frac{\lambda}{\tau}
\end{equation}

\noindent where
$\nu\frac{\partial u(0)}{\partial y}=U(-\frac{\cos
\omega t}{\delta_{-}}+\frac{\sin \omega t}{\delta_{+}})$ and

\begin{equation}
\sigma(0,t)=\rho \nu\frac{U}{1+\omega^{2}\tau^{2}}[-\frac{1}{\delta_{-}}(\cos
\omega t+\omega\tau\sin \omega t)+\frac{1}{\delta_{+}}(-\omega\tau\cos\omega
t +\sin \omega t)]
\end{equation}

\noindent  where $\rho$ is the density of a fluid. 
To obtain the dissipation rate per unit time per unit area of the plate, we calculate
$W=-\overline{u(0,t)\sigma(0,t)}$ averaged over a single cycle. The result is:

\begin{equation}
W(\tau,\omega)=\frac{1}{2} \frac{\mu  U^{2}}{1+\omega^{2}\tau^{2}}(\frac{1}{\delta_{-}}+\frac{\omega\tau}{\delta_{+}})
\end{equation}
\noindent The plot of the normalized dissipation rate as a function of $\omega\tau$ is shown
on Fig.1 .

\begin{figure}
\centerline{\includegraphics[angle=0,scale=1., ,draft=false]{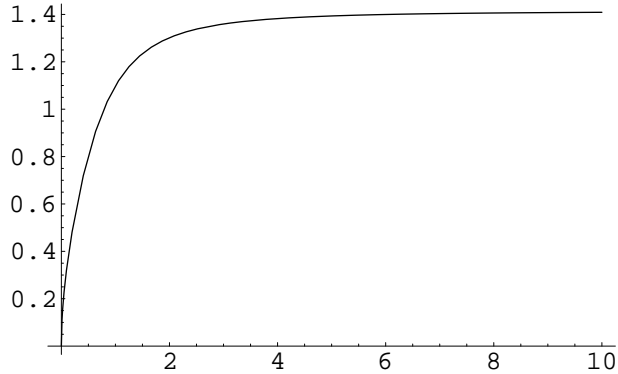}}
 \caption{  Dissipation rate $2W(\tau,\omega)$  as a function of $\omega$ (expression (16);  
(arbitrary units).}
\label{fig1}
\end{figure}

\noindent As $\omega\tau\rightarrow 0$, we recover the previous result $W/\sqrt{\omega}\propto
(1+\frac{\omega\tau}{2})$.  In the opposite  limit $\omega\tau\rightarrow\infty$,
the saturation of the curve $W(\omega\tau)$ is predicted.  In this range,  the kinetic energy of  the plate
oscillations is mainly dissipated into the wave radiation.  \\

\noindent {\it  "Plane oscillator".}
 Consider  a massless  spring of stiffness $k$ with two infinite plates of  height $h$ attached to it.  The losses in a spring are neglected and  the  friction force,  acting on the plates,  is the only source of  energy  dissipation. The equation of motion of this "plane oscillator"  driven by a force $R(t)$ is:
 
\begin{equation}
 x_{tt}+\gamma x_{t}+kx=R(t)
\end{equation}

\noindent where 
 $\gamma  x_{t}=\gamma u(0,t)=2\rho \sigma(0,t)/m_{s}$
is the friction force acting on a unit mass  ($m_{s}=\rho_{p}h$ ) of a plate surface  and the factor  2 accounts for the force acting on both (top and bottom) surfaces of the plate.   If the resonance $I(\omega_{0})$ is sharp enough,  so that 
$\delta \omega/\omega_{0}\ll1$, then we can calculate the damping (friction) force acting on the oscillator using the theory developed above.
 In Newtonian  limit  (hereafter, we omit the subscript $0$ ) $\omega\tau\rightarrow 0$, the expression (36) gives $\gamma=\sqrt{\mu\rho \omega/(2 m_{s}^{2})}\propto \sqrt{\omega}$ and the quality factor (the definition is given  below)  $Q\approx \omega/\gamma\propto \sqrt{\omega}$.   In the opposite (kinetic)  limit $\omega\rightarrow\infty$ and $\tau=const$:
 $
\gamma\rightarrow \mu\omega\tau
\sqrt{\frac{\omega}{\nu}}/(m_{s}(1+\omega^{2}\tau^{2})^{\frac{3}{4}})\propto
\omega^{0}=const =O(\rho c/m_{s})
$

\noindent This relation can be  understood as follows.  We are interested in a somewhat unusual case with the rapidly varying  deterministic velocity field $u(0,t)$ and  the slow - varying  ("frozen") molecular motion.  Thus,  the  long -time-averaging,   giving $\overline{u}=0$,  is unphysical  and the stress on a plane   $y=0$ 
is calculated by averaging over  the  plate surface:
$
\sigma_{xy}=\overline{(u(0,t)+c)_{x}(u(0,t)+c)_{y}}\approx u(0,t)c_{s}
$,  where $c_{s}$ is the mean velocity ($y$-component) of  molecules colliding with solid surface. In this limit, the  tangential component of  the force acting on a unit surface of a solid (per/unit mass) is:
$F_{t}=\frac{\rho c_{s}}{\rho_{s}h}u(0,t)$
and  $\gamma=\frac{\rho c_{s}}{\rho_{s}h}$. 
The former  relation, derived here for a rapidly oscillating plate,  is similar to the one appearing in a problem of a piston  instantaneously accelerated  in an ideal gas.  
Thus, in this limit the inverse quality factor   $1/Q=\frac{\gamma}{\sqrt{\omega^{2}-\gamma^{2}}}\approx
\gamma/\omega\propto 1/\omega$.  For a gas of a given temperature $\rho\propto p$
and $c\propto \sqrt{\theta/m_{mol}}=const$,   we have
 $
 \frac{1}{Q}\propto p/\omega
$

\noindent    
To obtain  the relation valid in the entire range of frequency variation  we have from (31):

\begin{equation}
\gamma=2\frac{\rho\sqrt{\omega \nu}}{\rho_{p}h(1+\omega^{2}\tau_{p}^{2})^{\frac{3}{4}}}[(1+\omega\tau_{p})\cos(\frac{1}{2}tan^{-1}\omega\tau_{p})-(1-\omega\tau_{p})\sin(\frac{1}{2} tan^{-1}\omega\tau_{p})]
\end{equation}

\noindent Our goal now is to find the relaxation time $\tau_{p}$ responsible for relaxation to equilibrium 
in the immediate vicinity of the rapidly oscillating   solid plate. 
In a standard equilibrium  situation,  when $\lambda$  and $\tau\approx \lambda/c\ll 1/\omega$ are   the smallest length and time-scales  in the system,  the relaxation times in the bulk and in the  immediate vicinity of the surface are more or less equal.  In a  rapidly oscillating flow,  we are interested in,  this is not so. Indeed, even in the air at  normal conditions,  the bulk  $\tau\approx \lambda/c\approx 10^{-9}sec$ and in the modern microresonators, where $\omega\approx 10^{9}sec^{-1}$,  the inequality  $\omega\tau\ll 1$ is hardly satisfied. Moreover, in the low pressure devices where $10^{-3}\leq p\leq 10^{3}torr$ , the bulk relaxation time is huge, so that $\tau_{p}$ must be calculated from a theory taking into account strongly non-equilibrium effects. In the absence of such  theory,  we use the results of experimental data on  a driven microbeam  obtained by Ekinci and Karabacak [3]   who covered
 an extremely wide range of parameter variation:  $10^{0}\leq p\leq 10^{3}torr$, $10^{6}\leq \omega \leq 0.6\cdot 10^{9} sec^{-1}$, $h=2\cdot 10^{-5}cm$. Under the normal conditions for $\omega=0.6\cdot 10^{9}sec^{-1}$ , the observed  damping parameter was  $\gamma\approx 1.5\cdot 10^{6}sec^{-1}$.
(This relation is the result of the measured $1/Q=\gamma/\omega\approx 3\cdot 10^{-3}$ and $\omega_{0}=0.6\cdot 10^{9}sec^{-1}$.)  Denoting  $x=\omega \tau_{p}$  and introducing  a geometric factor $g$ accounting for the geometry differences between theoretical and experimental setups, we have:

\begin{equation}
\frac{g}{4(1+x^{2})^{\frac{3}{4}}}[(1+x)\cos(\frac{1}{2}tan^{-1}x)-(1-x)\sin(\frac{1}{2} tan^{-1}x)]=1
\end{equation}

\noindent  In this paper we considered a case of the simplest possible geometry. In the nano-technological applications the resonator amplitude  is $O(10^{-6}cm)$ is much smallest than the smallest linear dimension of the beam (cantelever). Thus, the results presented here may be not too geometry -sensitive and can be readily generalized to different geometries.  
 For example,  the solution to the problem of an infinite cylinder  oscillating along its axis can be expressed in terms of the Bessel functions leading to $O(1)$  variations of the coefficient in front of (40).  Here, to qualitatively  illustrate the origins of the  geometric factor $g$,  we use very simple considerations.   As was stated above,  mass per unit area  of a plate  of height $h$ and length $l$ is : $m_{s}=\rho_{p}l^{2}h/l^{2}=\rho_{p}h$.  A simple calculation of $m_{s}$ for circular cylinder gives:   $m_{s}=\rho a/2$, where $a$ is the radius. In this case, $g_{cyl}\approx 4d_{plate}$. For a beam of height $h$ and width $w$, we have $m_{s}=\rho_{p}\frac{wh}{2(w+h)}$. Thus, the dissipation per unit area may be a relatively universal property [10]. 
The  equation (19), written for a particular magnitude of frequency ($\omega=0.6\cdot 10^{9}sec^{-1}$) and other parameters of experiment of Refs.[3], [10])  has  solutions for all $g\geq 3.6$.  For example,    $g=4$ and $g=6.28$ give $x\approx 1.8$ and $x\approx 4.8$  corresponding to  $\tau\approx 2 nsec $ and $\tau\approx  8nsec$, respectively.   Choosing $g=4$  we,  using  the  experimental data for normal pressure $p=800torr$ as a calibration point  take  $\tau_{p}(800torr)\approx 2nsec$  [3].  
Then,  based  on   kinetic theory,  
we substitute $\tau_{p}=\frac{1600}{p} nsec$ into  (39) and obtain $\gamma$ and quality factor $Q(\omega,p)$ in a good agreement with the results of Ekinci and Karabacak [3]  in a wide  range of both frequency and pressure variation ($10^{6}\leq \omega\leq 0.6\times 10^{9}Hz; ~~100\leq p\leq 1000 torr$. )

\noindent In  a  kinetic limit $\gamma=const$,   and introducing $G(x)=
[(1+x)\cos(\frac{1}{2} \tan^{-1} x)-(1-x)\sin(\frac{1}{2}\ tan^{-1} x)]/(1+x^{2})^{\frac{3}{4}}$,
the relation  (39) give:
$
a\sqrt{\omega}G(x)=1$, where the coefficient $a=a(p,\theta,m_{s},\rho,g)=const$ includes all   parameters of the system.  It is easy to check, that solution to this equation  $x\propto \omega$ defines an asymptotic kinetic time $\tau_{p}=const \neq \tau$ characterizing relaxation to equilibrium  in  the high-frequency limit $\omega\tau\rightarrow\infty$. Thus, the equilibrium  and non-equilibrium (high-frequency) relaxation times  may differ by a constant  $O(1)$ factor only. The inverse quality factor $1/Q(\omega)\approx \gamma/\omega$,  given by (39),  is shown on Fig. 2.  The most detailed comparison of the results based on relation (39) with experimental data is presented in Ref. 10.\\

\begin{figure} [h]
\centerline{\includegraphics[angle=0,scale=1., ,draft=false]{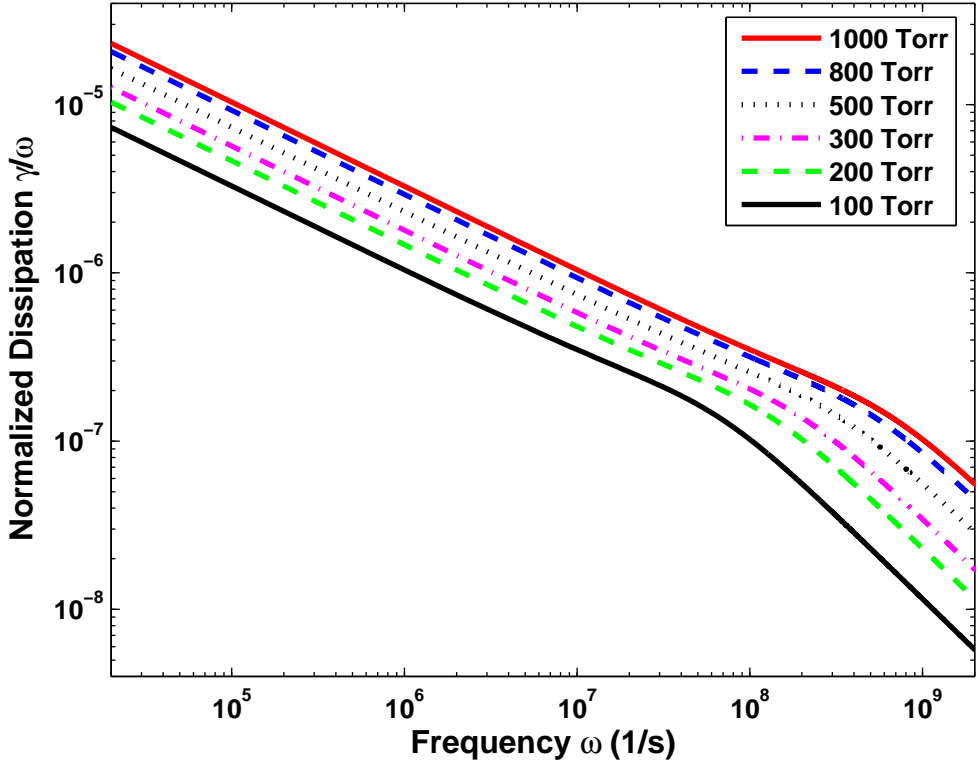}}
 \caption{  Inverse quality factor $1/Q\approx \gamma/\omega=\frac{4 \rho\sqrt{\nu}}{\rho_{p}h\sqrt{\omega}}G(x)\approx 60\sqrt{\frac{p}{800 \omega}}G(x)$  (relation (18)) vs  frequency $\omega$ in the presure interval $100<p<1000tor$.  Relaxation  time $\tau_{p}(800torr)=2 nsec$. For  remaining parameters, see text. }
\label{fig4}
\end{figure}

\noindent In a  dense liquid where $\omega\tau\ll1$, the relation (39) gives:

\begin{equation}
\gamma=\frac{\rho\sqrt{\omega \nu}}{\rho_{p}\frac{wh}{2(w+h)}}
\end{equation}
\noindent and in  water ($\nu\approx 0.01\frac{cm^{2}}{sec}$, $\rho/\rho_{p}\approx 2$,  $\omega\approx 0.6\cdot 10^{9} sec^{-1}$; $h=2\cdot 10^{-5}cm $ and $w=7\cdot 10^{-5}cm$), corresponding to the experimental set up  of Ref. 11,  we predict  $\gamma\approx 1.6\cdot 10^{8}sec^{-1}$ and $Q\approx  \frac{\sqrt{\omega^{2}-\gamma^{2}}}{\gamma}\approx 4$,  in a good agreement with the data [11].


  
\noindent {\it Summary and conclusions.}  In this paper  a  theory of  a flow generated by the oscillating plate,   valid in  a wide range of variation of dimensionless frequency  (Weissenberg number) $0\leq \omega\tau \leq \infty$, is  presented.  The solution to  kinetic    BGK equation,   describing transition between  viscoelastic and purely elastic regimes,  has been  found.  
The results obtained for a simple  "plane oscillator",  first introduced in this paper,  were favorably compared with the outcome of experiments on  nanoresonators of Refs. [3],[10] where this transtion has experimentally been  observed.

\noindent We would like to thank  K. Ekinci and D. Karabacak for sharing with us   their yet unpublished data on microresonators.  We benefited from the  most interesting and stimulating discussions with them and  R. Benzi, H. Chen, R. Zhang, X. Shan,  F. Alexander, S. Succi, I. Karlin and V. Steinberg.


\begin{thebibliography}{}


\bibitem{NonN1} R. B. Service,  {\em Tipping Scales-Just Barely}, Science {\bf 312}, 683 (2006);~
 H. G. Craighead, Science {\bf 290}, 1532 (2000); G. Binning, C.F. Quate, C. Gerber, Phys. Rev. Lett. {\bf 56}, 930-933 (1986).

\bibitem{NonN2} A.N. Cleland, M.L. Roukes, Nature {\bf 392}, 160 (1998);~X.M.Huang, C.A. Zorman, M. Mehregani and M.L. Roukes,   Nature, {\bf 421} 496 (2003); ~ K.L. Ekinci,  X.M.H. Huang and M.L. Roukes,  Appl. phys. Lett. {\bf 84}, 4469 (2004). 

\bibitem{NonN3} K.L. Ekinci and D. Karabacak, private communication (2006).

\bibitem{NonN4}  Xiaowen  Shan   H. Chen and R. Zhang, {\em Decay of the Shear Layer},  Phys. Rev. E (in press) (2006).

 \bibitem{NonN5} 
 G. G. Stokes, "On the effect of the internal friction in fluids on the motion of pendulums.", Cambr. Phil. Trans., {\bf IX}, 8 (1851);  L..D.  Landau and E.M. Lifshitz, {\em  Fluid mechanics},  
Pergamon Press,  Oxford (1959).

\bibitem{NonN6} H. Chen, S. A. Orszag, I. Staroselsky, and S. Succi,  {\em Expanded analogy between Boltzmann kinetic theory of fluids and turbulence},  J. Fluid Mechanics, {\bf 519} 301 (2004)
 
 \bibitem{NonN7} P.L. Bhatnagar, E.  Gross, and  M. Krook, \  Phys. Rev., {\bf 94}, 511--525 (1954);~ 
S. Chen and G. Doolen. Ann. Rev. Fluid Mech. 
{\bf 30}, 329 (1998);  H. Chen, S.  Kandasamy, S. Orszag,  R. Shock,
S. Succi, and V. Yakhot, 
Science {\bf 301}, 633--636, (2003).
 
\bibitem{NonN8} Landau, L. D. and Lifshitz, E. M. (1995): {\em Physical
Kinetics}, Butterworth/Heinemann.

\bibitem{Non9} C. Colosqui and V. Yakhot,  "Lattice Boltzmann Simulations of Second Stokes Flow Problem", J. Modern Physics (in press); F. Alexander 2006, (private communication).


\bibitem{Non9}  D. M. Karabacak,  V. Yakhot and K. Ekinci, "High frequency nanofluidics: experimental study with nano-mechanical resonators", ArXiv, cond-mat/0703230.
 
\bibitem{Non10}  S.S. Verbridge, L.M. Bellan, J.M. Parpia and H.G. Craighead, Nano Lett, {\bf 6}, 2109 (2006).
.





\end{thebibliography}
\end{document}